\documentstyle[prl,aps,epsf,epsfig,multicol,psfrag]{revtex}

\newcommand{\be}{\begin{equation}}
\newcommand{\ee}{\end{equation}}
\newcommand{\ba}{\begin{array}}
\newcommand{\ea}{\end{array}}

\begin{document}

\title{Stretched Polymers in a Poor Solvent}

\author{Peter Grassberger and Hsiao-Ping Hsu}

\address{John-von-Neumann Institute for Computing, Forschungszentrum J\"ulich,
D-52425 J\"ulich, Germany}

\date{\today}

\maketitle

\begin{abstract}
Stretched polymers with attractive interaction are studied
in two and three dimensions. They are described by 
biased self-avoiding random walks with nearest
neighbour attraction. The bias corresponds to opposite forces 
applied to the first and last monomers.
We show that both in $d=2$ and $d=3$ a phase transition occurs 
as this force is increased beyond a critical value, where the 
polymer changes from a collapsed globule to a stretched configuration.
This transition is second order in $d=2$ and first order in $d=3$.
For $d=2$ we predict the transition point quantitatively
from properties of the unstretched polymer. This is not possible
in $d=3$, but even there we can estimate the transition point
precisely, and we can study the scaling at temperatures slightly 
below the collapse temperature of the unstretched polymer. We 
find very large finite size corrections which would make 
very difficult the estimate of the transition point from 
straightforward simulations.
\end{abstract}

\begin{multicols}{2}

\section{Introduction}

Deformations of polymers have been studied for many years, such
as stretching a single DNA or titin macromolecule 
\cite{per94,per95,bus94,kell97,oest97,tskh97}. Such experiments will 
become even more important with the rapid improvement in
single molecule experiments \cite{phystoday}. Their theoretical
understanding has attracted much attention. Most such experiments have been 
performed in good solvents, where also the theory is best understood
\cite{deGennes}.
But of particular interest in biology is the stretching of a collapsed polymer, 
i.e. of a polymer in a poor solvent below the $\Theta$-temperature.
Up to now there seem to exist only few experiments in this regime 
\cite{haupt}. Unfolding proteins in this 
way could  e.g. give important information on their spontaneous 
folding pathways. 

Although there exist several theoretical papers on stretching of collapsed 
polymers \cite{lai95,lai96,halp91a,halp91b,lai98,wittkop,wittkop95},
we believe that further work is needed for a full understanding. It is
generally believed that there is a first order transition between the 
globule and the stretched phases in 3 dimensions. But it seems that 
precise estimates of the critical force do not exist. Also, there are 
no predictions for the scaling laws expected when the temperature 
approaches the collapse temperature $T_\theta$ from below. Finally, 
it seems that there is only one paper \cite{wittkop95} which deals with
$d=2$. Moreover, while it is claimed in \cite{wittkop95} that the transition 
is also first order in $d=2$, we shall find there a rather different 
situation and a second order transition.

A stretched polymer in a bad solvent is modeled as a biased interacting 
self-avoiding random walk (BISAW) on a regular lattice (square in $d=2$, 
simple cubic in $d=3$) with nearest neighbour attraction. In this model no two
monomers can visit the same site. The attraction is taken into account 
by a Boltzmann factor $q=e^{-\beta \epsilon}$ for each pair of non-bonded 
monomers occupying nearest neighbours on the lattice. Here $\beta=1/kT$ and 
$\epsilon < 0$ is the attractive potential between non-bonded nearest-neighbour
pairs. The stretching is described by a factor $b = \exp(\beta a F)$ where 
$a$ is the lattice constant and $F$ is the stretching force~\cite{footnote}.
The partition sum is therefore
\be
   Z=\sum_{walks}q^{m}\, b^{x}.                    \label{SAW}
\ee
where $m$ is the number of non-bonded occupied nearest-neighbour pairs,
and $x$ is the distance (in units of lattice constants) between the two
end points of the chain in the direction of $\bf F$. We are interested in
polymers which form collapsed globules if they are not stretched, i.e. 
$q > q_\theta$ where $q_\theta = e^{-\epsilon/kT_\theta}$.


In the present paper we employ the Pruned-Enriched-Rosenbluth Method (PERM)
\cite{g97} to study the phase transition of the BISAW in $d=2$ and 3. 
But we shall see that, in order to understand the BISAW, we need also some 
other results. More precisely, we will need also some more results on the
unbiased interacting self-avoiding random walk (ISAW),
both on infinite and on finite lattices.
On finite lattices with periodic boundary conditions we can study the 
bulk behaviour of collapsed polymeric matter without being disturbed by 
surface effects \cite{gh95,g97}.

As we shall see, a stretched collapsed polymer in $d=2$ forms, in the 
infinite chain length limit $N\to\infty$ and for forces below the transition
to the stretched phase, a compact object of a shape shown in Fig.~1.
There the boundaries are circular arcs, the density inside is independent 
of the stretching force $F$, and the shape (i.e. the angles at the two 
extreme points) is determined both by $F$ and by $q$. The transition 
to the stretched phase occurs when these angles tend to zero. 

In $d=3$ the 
situation is more complicated because of the Rayleigh instability. For 
$F<F_c$ one has an object of roughly elongated ellipsoidal shape, 
but for $F=F_c$ two phases coexist: Part of the chain is stretched, 
while the rest is still collapsed (Fig.~2). In the collapsed part, the 
density is the same as in an unstretched globule (for $N\to\infty$). 
The stretched part might consist of a single piece ({\it tadpole}
configuration \cite{ball}), or of two pieces as in Fig.~2. Within our 
theory both configurations have the same energies. The critical force 
$F_c$ vanishes linearly when $T \to T_\theta$.

The paper is organized as follows: In Sec.~II we treat the 2-dimensional
case. We first formulate in more detail the model sketched above, and we 
show that it is in good agreement with simulations both of stretched and 
unstretched chains. The 3-dimensional case are discussed in Sec.~III.
Conclusions are drawn in Sec.~IV. Finally, details of the Monte Carlo algorithm
are given in an appendix.

\begin{figure}
  \begin{center}
   \psfig{file=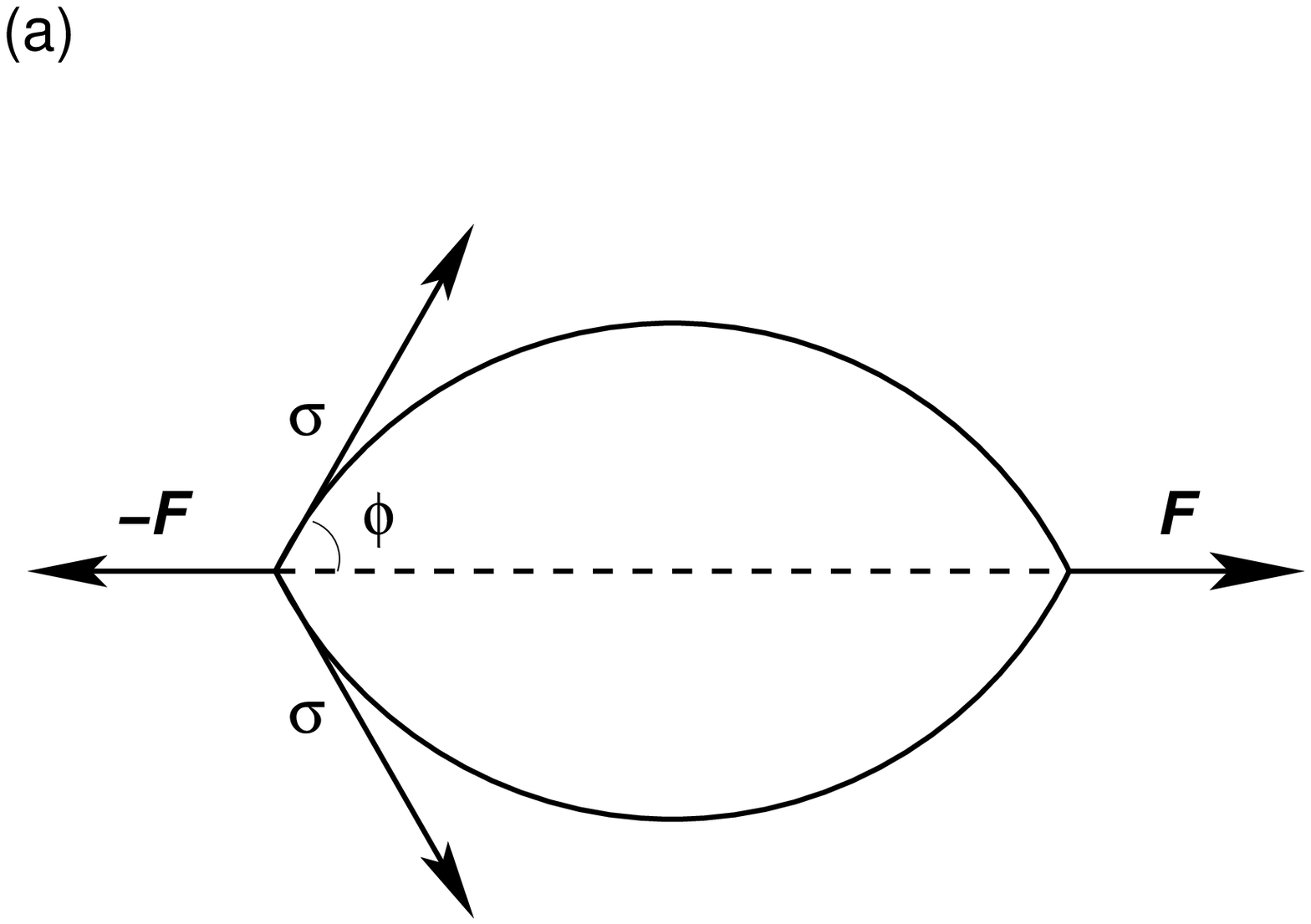,width=5.4cm, angle=0}
   \vskip 0.5 true cm
   \psfig{file=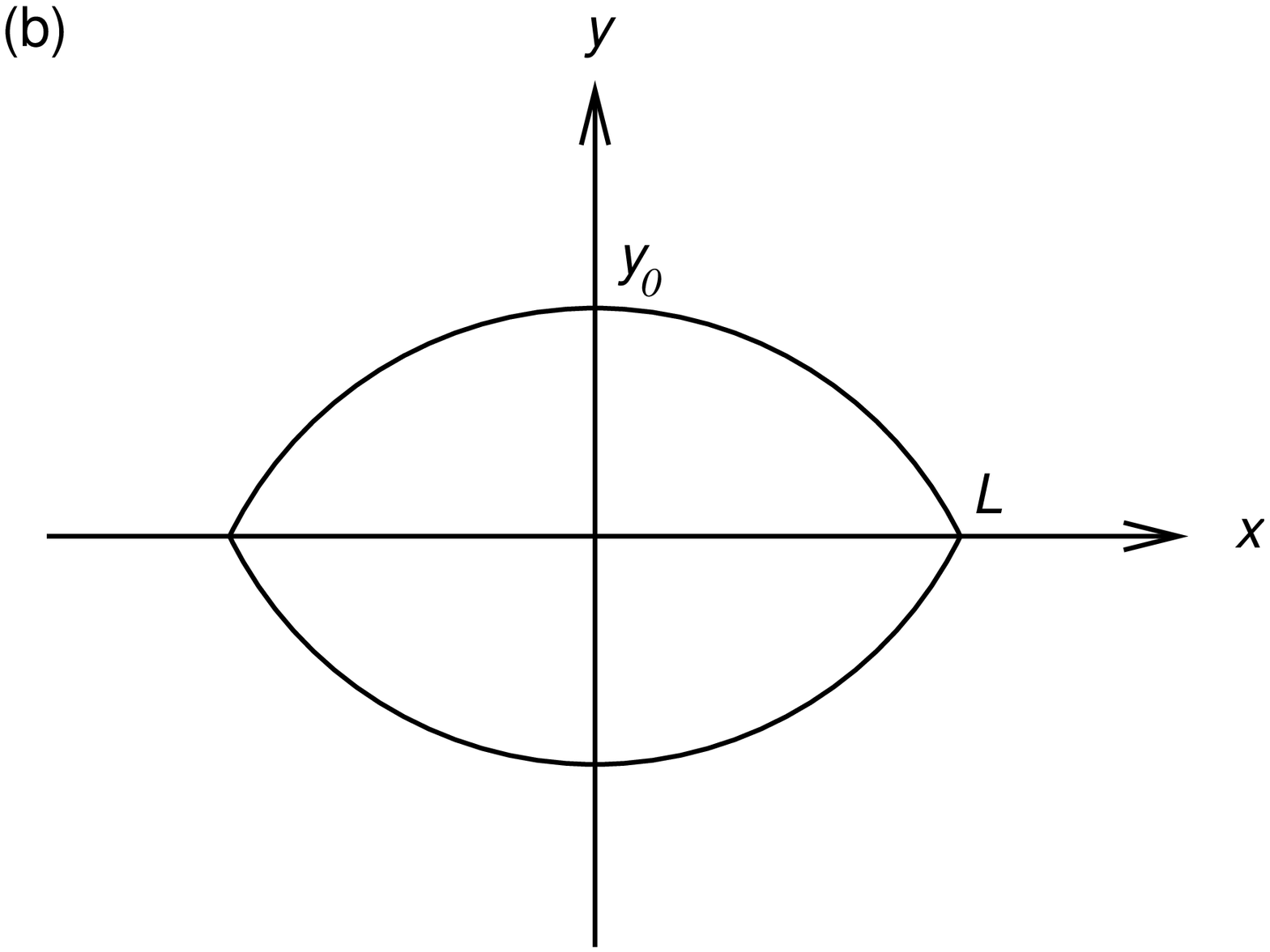,width=5.4cm, angle=0}
   \vskip 0.3 true cm
   \label{globule2d}
   \caption{(a) The geometric shape of a stretched polymer in a poor
             solvent in $d=2$, 
            (b) Coordinates used in the calculation.}
  \end{center}
\end{figure}

\begin{figure}
  \begin{center}
   \psfig{file=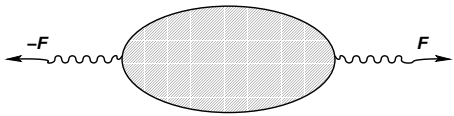,width=6.0cm, angle=0}
   \vskip 0.3 true cm
   \label{globule3d}
   \caption{Schematic drawing of a stretched polymer in a poor solvent in $d=3$, 
       at an intermediate force {\bf F} where the collapsed and extended 
       phases coexist.}
  \end{center}
\end{figure}

\section{D=2}

Let us first consider an unstretched polymer in $d$ dimensions, 
for any $d\ge 2$,
below the $\Theta$-point. In the following, the lattice constant will always be 
$a=1$. For large $N$ and for infinite lattices the polymer forms a 
{\it globule} with a bulk monomer density $\rho(T)$ which tends
to zero for $T \to T_\theta$, $\rho(T) \sim (T_\theta-T)^\beta$. For $d\ge 3$ we
have $\beta=1$ and logarithmic corrections \cite{g97}, but for $d=2$ the 
power $\beta$ seems to be unknown. For $T\to 0$ we have $\rho(T)\to 1$.

The free energy for such a polymer consists of two parts: An extensive bulk
contribution $\propto N$, and a surface contribution $\propto N^{(d-1)/d}$.
For $d=2$ this reads
\be 
   - \log Z_N(q, b=1) \approx \mu_\infty(q) N + \tilde{\sigma}(q) \sqrt{N}
                   \label{Z2N}
\ee
with $\mu_\infty$ being the chemical potential per monomer in an infinite chain,
and $\tilde{\sigma}$ is related to the surface tension (=free energy per 
unit of perimeter length) $\sigma$ by
\be
    \tilde{\sigma} \sqrt{N} = 2\pi R \sigma
\ee
with $\pi R^2 \rho = N$. Combining the last two equations we obtain
\be 
    \sigma = {1\over 2}\tilde{\sigma}\sqrt{\rho\over \pi}\;.
                       \label{sigma}
\ee

Now assume we start stretching the polymer by pulling at the two ends.
The first effect will be that the ends are drawn toward the surface and 
the whole globule is rotated such as to minimize the potential energy.
We will neglect the change in entropy associated to that. When the 
stretching force increases, we develop a shape as indicated in Fig.~1a.
We shall assume that the bulk is incompressible, so that we still have 
the same bulk free energy, and we obtain our final ansatz
\be
   - \log Z_N(q, b) \approx \mu_\infty N + \sigma P - 2LF
\ee
where $P$ is the perimeter and $2L$ is the end-to-end distance.
Parameterizing the shape by a function
$y = y(x)$ with $y(0)=y_o$, $dy/dx|_{x=0} = 0$, and $y(L)=0$ (see Fig.~1b),
we have 
\be
   P = 4 \int_0^L dx \sqrt{1+(dy/dx)^2}, \qquad 4 \rho \int_0^L dx y(x) = N.
\ee

The optimal shape is obtained by maximizing $Z_N$ at constant $N$. The Euler
equations for this maximization problem consist of two parts. The first,
resulting from $\partial Z_N(q, b)/\partial L = 0$, gives the equilibrium
condition between stretching and surface tension forces,
\be
   F = 2 \sigma \cos\phi,
\ee
where the angle $\phi$ is indicated in Fig.~1a.
Thus the critical point, corresponding to $\phi = 0$, is at $F=2 \sigma$ or
\be
   b_c = e^{2 \sigma} \;.
                           \label{bcs}
\ee
The second part, maximization with respect to $y(x)$, gives that $y(x)$ is a circular
arc. Using the angle $\phi$ indicated in Fig.~1a, we have then
\be
   P = 4L{\phi\over \sin\phi}, \qquad N/\rho = 2L^2 {\phi-\sin\phi \cos\phi \over \sin^2\phi}.
\ee
For $b<b_c$ we have
\be
   - \log Z_N(q, b) = \mu_\infty N + \sigma N^{1/2}[8 (\phi-\sin\phi\cos\phi)/\rho]^{1/2} 
                          \label{Z2}
\ee
with
\be
   \phi = \arccos {\log b\over 2\sigma} \;.
                            \label{phi}
\ee

Eqs.~(\ref{bcs})-(\ref{phi}) form our final solution. They involve the three 
temperature dependent material constants $\mu_\infty$, $\sigma$, and $\rho$. The former 
two can be estimated from Eq.~(\ref{Z2N}) if we measure in addition the gyration radius, 
but more precise estimates of $\mu_\infty$ and $\rho$
result from simulations on finite lattices in the dense limit \cite{hegger,g97}.
Thus our strategy in verifying the above theory numerically consists of the 
following steps:
\begin{enumerate}
\item We simulate chains on finite lattices of size $L\times L$ (typically with $L=8$
to 64) by means of the PERM algorithm which gives us directly estimates
of the partition sum. For $q>q_\theta$ its logarithm is not convex. For suitably 
adjusted $\mu_L$, $\log Z_N(q, b=1, L) + \mu_L N$ has two peaks of equal heights: 
One at $N\approx 0$, the other at $N\approx L^2\rho$. Extrapolating to $L=\infty$ we 
obtain $\mu_\infty$.
\item In the next step we simulate chains on (practically) infinite lattices, again by means 
of PERM. Using Eq.~(\ref{Z2N}) and the already obtained values of $\mu_\infty$ and 
$\rho$ we then obtain $\sigma$.
\item Finally we simulate stretched polymers, to compare with the prediction
of Eq.~(\ref{Z2}). 
In addition to $Z_N(q, b)$ we measure in these runs also $\langle x \rangle$
which should be equal to $2L$ for large $N$
as long as $b<b_c$. We should, however, immediately warn 
that these latter measurements are not very conclusive, since finite size corrections 
are large in this regime. We might point out that for $b=1$ there is the 
exact result 
\be
   b{\partial \over \partial b} \langle x \rangle = \langle x^2\rangle.
\ee
Since $\langle x^2\rangle \sim N$ for collapsed polymer in $d=2$, we thus have
$\langle x \rangle
\sim N$ for very small $b$. This shows that the above model (which would give 
$\langle x \rangle \sim N^{1/2}$) cannot be correct for small $N$, as we had indeed
pointed out already before.
\end{enumerate}

In Fig.~3 we show $\log Z_N+ \mu_L N $ for $q=2.4$, $b=1$, and for finite lattices with 
$L^2 = 2^7,\; 2^8,\; \ldots,\; 2^{12}$ sites, plotted against $N/L^2$. The curves are in
the same order as shown in the legend.
Since $q_\theta \approx 1.95$
for this model \cite{barkema}, this is deep inside the collapsed regime. We used helical 
boundary conditions. When $L^2$ was not an integer, the 
effective lattice shape was not a perfect square, but this led to negligible corrections. 
The values of $\mu_L N $ were fixed by demanding the right hand peak to have height
exactly zero. The values obtained this way are shown in Fig.~4. They are plotted there
against $L^{-1.72}$ because this gave the best straight extrapolation to $L=\infty$.
The extrapolated value is $\mu_\infty = -1.3213(1)$. The positions of the right hand peaks
give finite lattice approximations to the density $\rho$. Extrapolating to $L=\infty$
gave $\rho = 0.84(1)$. Analogous simulations were also done at different values of $q$.
   
\begin{figure}
  \begin{center}
   \psfig{file=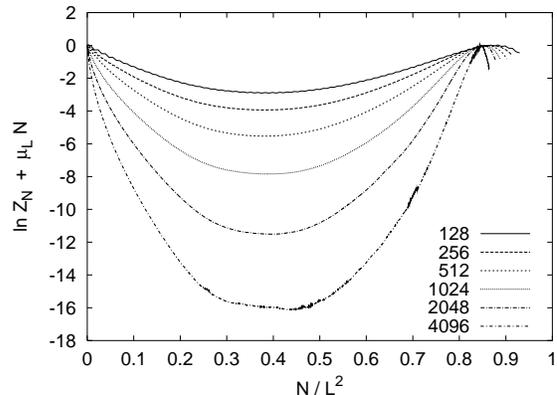,width=5.4cm, angle=270}
   \label{logzn3}
   \caption{For $d=2$, $q=2.4$ and $b=1$, $\ln Z_N + \mu_L N$ versus $N/L^2$ for
       finite lattices with $L^2=2^7,\; 2^8,\; \ldots, \; 2^{12}$. The values 
       of $\mu_L$ were fixed by demanding that the peaks at $N/L^2\approx 0.85$ 
       have zero height.}
  \end{center}
\end{figure}

\begin{figure}
  \begin{center}
   \psfig{file=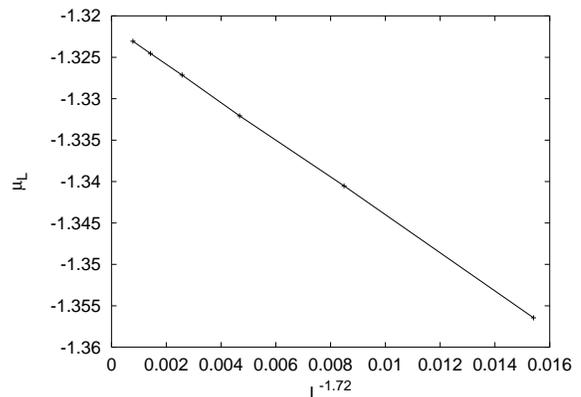,width=5.4cm, angle=270}
   \label{muinfty}
   \caption{$\mu_L$ as obtained from Fig.~3, plotted against $L^{-1.72}$ which 
      gave the most straight extrapolation. Error bars are ca. 0.00005, smaller 
      than the symbol sizes. The extrapolation to $L=\infty$ is $\mu_\infty = -1.3213
      \pm 0.0001$.}
  \end{center}
\end{figure}

Results from unbiased ISAW on an `infinite' lattice (i.e. on a lattice that
was so large that no walk reached the boundary) are shown in Fig.~5. There we
plot $\log Z_N + \mu_\infty N$ against $N^{1/2}$. From Eq.~(\ref{Z2N})
we expect this to give a straight line with slope $-\tilde{\sigma}$.
Actually the line is slightly curved, indicating that there
are further finite size corrections for small $N$ and systematic
sampling corrections at large $N$. The latter are indeed to be expected.
When we constructed histograms of tour weights (see the appendix), 
we found that the simulations are unreliable for $N>1000$.
We should have shown only the data for $N<1000$, 
but we showed all data, for the following reasons: 
\begin{itemize}
\item we want to compare these simulations with biased and finite
volume simulations
of the same length which are easier for PERM and which are thus still reliable
for $N=3500$, the longest chains used in Fig.~5.
\item Even if we cannot be sure 
that the data for $N>1000$ are correct, we cannot argue either
that they must be wrong. They are most likely too low, since PERM has
difficulties to sample configurations which start out (as the chain grows)
to have high energy (i.e. few contacts), but which become `good' 
during later growth stages. But we do not know how important this is. 
\end{itemize}

The curvature observed in Fig.~5 dominates the error in our estimate 
$\tilde{\sigma} = 0.46(2)$. Combining this with the previous estimate for 
$\rho$ we obtain $\sigma = 0.119(5)$, and from this we predict $b_c = 1.269(13)$.
Remember that all this is for $q=2.4$.

Finally, results from simulations with $b\geq 1$ are shown in Figs.~6 to 9. In Fig.~6 we 
show $\log Z_N(q,b) + \mu_\infty(q) N$ versus $N$. We expect these curves to become 
horizontal for $N\to\infty$ as long as $b\leq b_c$, while they should 
increase linearly for $b>b_c$. This is indeed seen, although we now find
a slightly larger value $b_c = 1.285(6)$. In spite of the small discrepancy 
with the predicted value we consider this as remarkable agreement. Our data are 
not precise enough to allow a comparison with the detailed predictions for 
$1<b<b_c$. 

For $b>b_c$
the chemical potentials $\mu^{(s)}_\infty(q,b)$ in the stretched phase can be
estimated easily by demanding that $\log Z_N(q,b) + \mu^{(s)}_\infty(q,b) N$ 
becomes $N$-independent for large $N$. The results 
are shown in Fig.~7. Although this figure clearly shows a continuous transition, 
the data are not precise enough to quote a meaningful critical exponent. It 
is compatible with $3/2$, but the uncertainty is large.

Among the other measurements during these runs, the most interesting are those 
of $\langle x\rangle$. Again they are not precise enough for a detailed 
comparison with the predictions in the regime $b<b_c$. But they also show
clearly the phase transition at $b=b_c$, since $\langle x\rangle$ rises 
linearly with $N$ only in the stretched phase (see Fig.~8).
Values of $v = d\langle x\rangle/dN$,
obtained by extrapolating the observed slopes in plots of $\langle x\rangle$
against $N$ toward $N=\infty$, are shown in Fig.~9. Again we see a continuous
transition with an exponent which is roughly equal to $0.7$, but uncertainties
are too large to make a more definite statement.

\begin{figure}
 \begin{center}
   \psfig{file=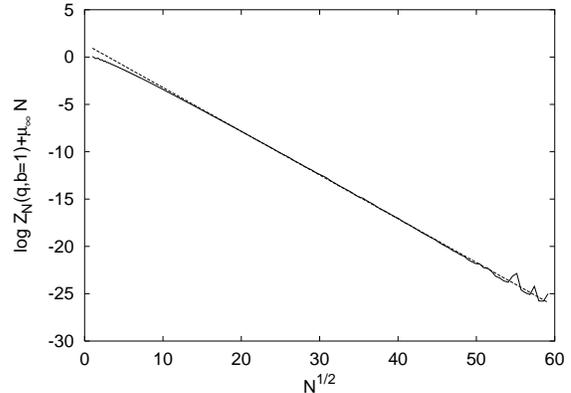,width=5.4cm, angle=270}
   \label{zg-2d}
   \caption{Values of $\ln Z_N(q,b=1) + \mu_\infty N$ versus $N^{1/2}$ for $d=2$,
     $q=2.4$, and $\mu_\infty=-1.3213$. The dashed line has
     slope $-\tilde{\sigma}$ $(\tilde{\sigma}=0.46 \pm 0.02)$.}
  \end{center}
\end{figure}

\begin{figure}
 \begin{center}
   \psfig{file=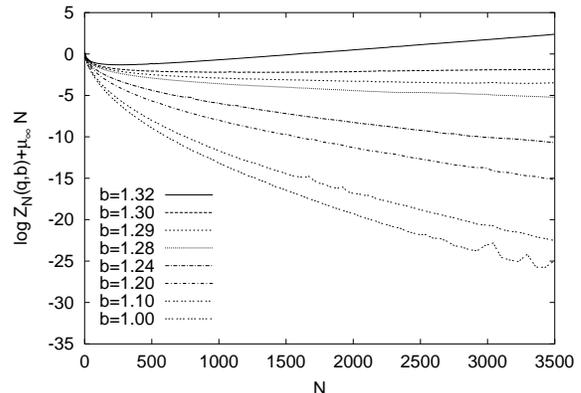,width=5.4cm, angle=270}
   \label{zg-2db}
   \caption{Values of $\ln Z_N(q,b) + \mu_\infty N$ versus $N$ for the same 
     $q$ and $\mu_\infty$ as in Fig.~5, and for various values of $b$.}
  \end{center}
\end{figure}

\begin{figure}
 \begin{center}
   \psfig{file=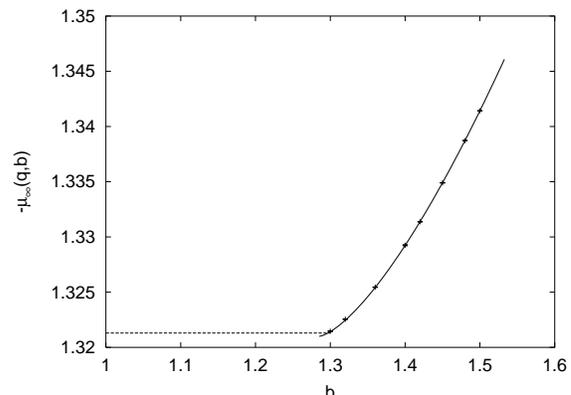,width=5.4cm, angle=270}
   \label{zg-2dc}
   \caption{Values of $\lim_{N\to\infty}[N^{-1}\ln Z_N(q,b)]$ versus $b$,
     for $d=2$ and $q=2.4$.
     For $b>b_c$ this is $-\mu^{(s)}_\infty(q,b)$.
     For $b<b_c\;(\approx 1.3)$, we replaced the actual numerical estimates
     by the theoretical value $-\mu_\infty(q)$.}
  \end{center}
\end{figure}

\begin{figure}
 \begin{center}
   \psfig{file=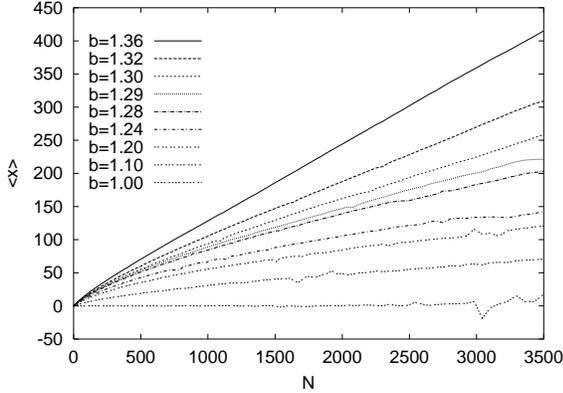,width=5.4cm, angle=270}
   \label{x-2d}
   \caption{Average displacement $\langle x \rangle$ in the bias direction
            for $d=2,\; q=2.4$ and for various values of $b$.}
 \end{center}
\end{figure}

\begin{figure}
 \begin{center}
   \psfig{file=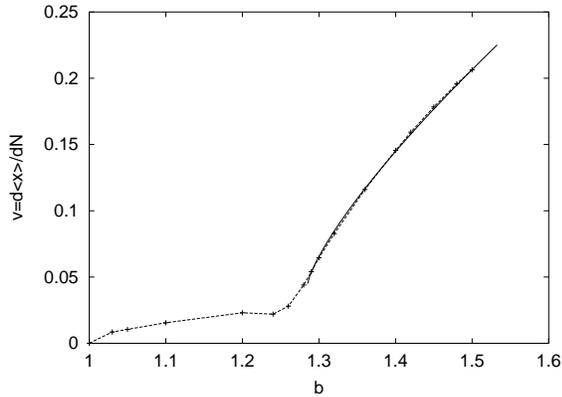,width=5.4cm, angle=270}
   \label{v-2d}
   \caption{Asymptotic ``velocity" $d\langle x \rangle/dN$ in the bias
       direction for $d=2,\; q=2.4$ and for various values of $b$.}
  \end{center}
\end{figure}

Before leaving this section we should remark that we checked carefully,
in view of Ref.\cite{wittkop95}, that the unfolding transition is indeed 
second order in our model. We looked e.g. at histograms analogous to those 
in Figs.~12 and 13. We definitely saw no hint of bistability or hysteresis.
Together with the good agreement with the above model this seems definitely
to rule out a first order transition. We cannot rule out, however, that 
the transition is first order in another microscopic model such as the 
bond fluctuation model studied in \cite{wittkop95}. 

\section{D = 3}

Some of the treatment for $d=2$ can be carried over to 3 dimensions with minor 
modifications. Instead of Eq.~(\ref{Z2N}) we now have \cite{owcz,gh95}
\be
   - \log Z_N(q, b=1) \approx \mu_\infty(q) N + \tilde{\sigma}(q) N^{2/3}\;.
                   \label{Z3glob}
\ee
Again we estimate $\mu_\infty$ most reliably from finite lattice simulations in 
the dense limit \cite{g97}. Again $\tilde{\sigma}$ can be related to a 
surface tension, and can be estimated numerically by plotting 
$\log Z_N(q, b=1) + \mu_\infty(q) N $ against $N^{2/3}$.
  
In the weak stretching regime we could still try to solve the optimal shape,
but this time the minimization problem is more complicated and we were not 
able to find an explicit solution. But more important, we do not expect a 
continuous transition as in $d=2$. The reason is that we expect a Rayleigh 
instability when the globule is stretched too much, which is expected to 
occur, before the limit $\phi=0$ is reached. Thus we cannot expect to be
able to predict the transition point as in $d=2$.

On the other hand, things also simplify because the critical exponents at 
the $\Theta$ point now should be mean field like (the upper critical 
dimension for the $\Theta$ point is 3 \cite{deGennes}). This means in
particular that $\tilde{\sigma}, \rho$ and $\mu_\infty(q) - \mu_\infty(q_c)$ 
all should vanish as $\sim (T_\theta - T)$ when $T\to T_\theta$ from below.

For a stretched globule with $b<b_c$ we expect that $Z_N$ differs from the 
value for $b=1$ only by its changed surface, i.e. by terms $\sim N^{2/3}$.
For $b>b_c$ we have a string-like phase whose free energy should increase 
linearly with $N$ with very small corrections, 
\be
   -\log Z_N(q, b>b_c) \approx \mu^{(s)}(q,b) N \;,
                    \label{Zstretch}
\ee
where the chemical potential $\mu^{(s)}(q,b)$ in the stretched phase is a 
function independent of the chemical potential $\mu_\infty(q,b)$ in the 
globular phase. The end-to-end distance in the stretched phase is given 
by $\langle x\rangle = N v(q,b)$ with
\be
   v(q,b) = - b {\partial \mu^{(s)}(q,b) \over \partial b}\;.
                \label{x-stretch}
\ee
Stability of the stretched phase requires that $\partial v(q,b) /
\partial b > 0$, i.e.
\be
   b {\partial \mu^{(s)}(q,b) \over \partial b},\;\; 
   (b {\partial \over \partial b})^2 \mu^{(s)}(q,b) < 0\;.
             \label{stable}
\ee
 
In between these two regimes we expect a coexistence region where part 
of the chain forms a (single) globule, while the rest forms one or two 
stretched pieces. 
A somewhat crude model of a polymer in the coexistence region 
which, however, catches all essential features including finite size 
effects, is the following: The stretched part has $N_s$ monomers, and
its partition sum is described by Eq.~(\ref{Zstretch}) with $N$ replaced 
simply by $N_s$. The globular part has $N_g = N-N_s$ monomers, and 
its partition sum is described by Eq.~(\ref{Z3glob}) with $N$ replaced 
by $N_g$. The total partition sum is just the product of the two, i.e.
the free energy is just the sum of free energies of the two parts.
Notice that this involves a number of approximations:
\begin{itemize}
\item There is no penalty for the area where the globule and the 
stretched part(s) are attached to each other. Such a free energy 
contribution should be independent of $N$ and can be safely neglected.
\item The globule is approximated by a sphere. This is a 
more serious approximation. It systematically overestimates the free 
energy, by an amount $\propto N_g^{2/3}$. This is of the same order
of magnitude as if a wrong surface tension were used. It should 
therefore lead to quantitative errors, but not to qualitative ones.
Moreover, the errors should be small because too elongated globules
are prevented by the Rayleigh instability.
\item We neglect all fluctuations.
\end{itemize}

The total free energy in the coexistence region is thus, for a fixed total 
end-to-end distance $x$ (and not denoting explicitly the dependence on $q$),
\be
   -\log Z_N = N_g\mu_\infty + N_g^{2/3}\tilde{\sigma} +(N-N_g) \mu^{(s)}(b)
              - x F \; .
\ee
Minimizing this with respect to $N_g$ we obtain 

\begin{eqnarray}
   \mu_\infty - \mu^{(s)}(b) + {2\tilde{\sigma} \over 3} N_g^{-1/3} =0 \;.
                          \label{long}
\end{eqnarray}
This can be read as an implicit equation for $b$. If solved, it gives $b$ as 
a function of $N_g$. Notice that $b$ is independent of $N_s$:
According to our model the force needed to pull out 
the chain from the globule depends on its size through the surface tension 
(last term in Eq.~(\ref{long})), but not on the length $N_s$ of the
stretched part.

In the limit $N_g\to\infty$ this term does not contribute, and we obtain 
the condition for the true transition point
\be
   \mu^{(s)}(b_c) = \mu_\infty \;. \label{mubc}
\ee
For finite $N_g$ we see that at the coexistence point $\mu^{(s)}(b)$ decreases 
with $N_g$. Since $\partial \mu^{(s)}(b)/\partial b < 0$, this implies that
the effective critical $b$ increases with $N_g$. In other words, as the 
chain is pulled out from the globule, $N_g$ shrinks, and thus the force needed
to pull out more of the chain decreases. This is the basic instability which 
makes the transition first order and shows that our ansatz is consistent. 
It implies that a finite globule will be entirely pulled open as soon as 
$b > b_c - const/N^{1/3}$. Notice that
a more realistic model (where the spherical globule is replaced by 
some other shape with limited aspect ratio) would still give the same 
qualitative results: For infinitely large globules, the critical point is 
given by equating $\mu^{(s)}(b_c)$ with the chemical potential for 
{\it unstretched} globules; and finite size corrections to this are negative
and decrease as $N^{-1/3}$.

In a first set of simulations we determined numerically $\mu_\infty(q)$ 
and $\rho(q)$ for a wide range of $q$. We did this again, as in $d=2$, by 
performing the simulations on finite lattices in the
dense limit. In a second step, we obtained $\tilde{\sigma}$ from simulations 
of unbiased ISAWs in infinite lattices. In contrast to $d=2$, here we 
encountered the problem that there are significant further corrections 
to the asymptotic ansatz Eq.~(\ref{Z3glob}). We illustrate this in Fig.~10
where we plot $\log [Z_{N-n}(q)/Z_{N+n}(q)]/(2n)$ against $N^{-1/3}$
for $b=1$ and several values of $q$. Here, $n = 1+\lfloor N/20 \rfloor$.
The points on the $y$-axis are obtained from dense limit simulations.
The straight lines are extrapolations constrained to pass through these
points. Their slopes are $2\tilde{\sigma}/3$. While the curves are
compatible with these extrapolations, close inspection shows that they
are not convex as one might have guessed naively. Thus, extrapolations
not aided by the dense limit simulations would be prone to large errors.

\begin{figure}
 \begin{center}
   \psfig{file=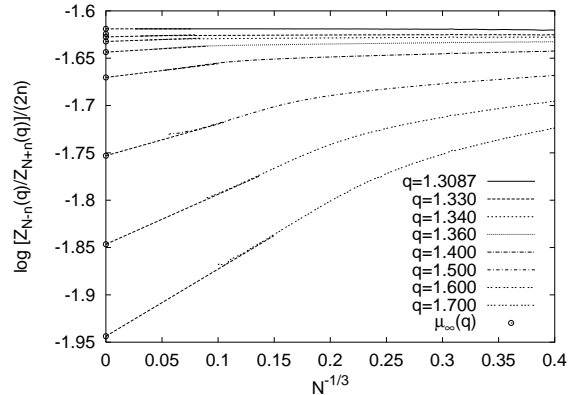,width=5.4cm, angle=270}
   \label{logz3d}
   \caption{$\ln [Z_{N-n}(q)/Z_{N+n}(q)]/(2n)$ versus $N^{-1/3}$
      for $d=3,\;b=1$ and several values of $q$. The points on the $y$-axis are
      obtained from dense limit simulations on finite lattices, and the
      straight lines are extrapolations to $N\to\infty$, constrained to
      pass through these points.}
  \end{center}
\end{figure}

Finally, we performed simulations of BISAW (on `infinite' lattices).
We shall discuss in detail only the most extensive simulations, done at
$q=1.5$ which is deep in the collapsed region ($q_\Theta=1.3087(3)$ for this
model \cite{g97}). But similar simulations were also done at
different values of $q$.

\begin{figure}
  \begin{center}
   \psfig{file=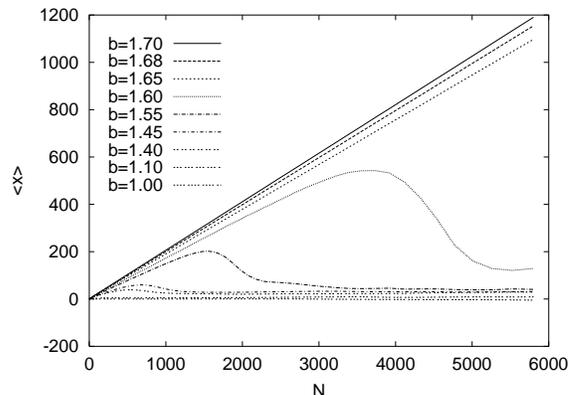,width=5.4cm, angle=270}
   \label{x-3d}
   \caption{Average displacement in the bias direction $\langle x \rangle$
    for $d=3,\; q=1.5$ and for various values of $b$.}
  \end{center}
\end{figure}

Average displacements $\langle x \rangle$ against $N$
for various biases $b$ are shown in Fig.~11. For $b=1$, we have of course
$\langle x \rangle = 0$. With increasing $b$, $\langle x \rangle$ still
remains close to zero for large $N$ and increases very slowly as $N\to\infty$
as long as $b$ is close to 1. But for small $N$ it increases roughly
$\propto N$, with an abrupt decrease in a narrow range which shifts toward
larger $N$ when $b$ is increased.
As $b$ increases above a certain value $b_c$, the linear range of
$\langle x \rangle$ versus $N$ extends to $N=\infty$, indicating the
phase transition from the collapsed to the stretched polymer phase.
From Fig.~11 the critical point of the finite system might be guessed
to be between $1.60$ and $1.65$,
but we will see that finite-size corrections are very large and $b_c$
is actually larger than $1.85$.

Very similar behaviour is seen for $R^2_N$ and for the average number of
non-bonded nearest-neighbour pairs $\langle m \rangle$ (not shown). For $b>b_c$
they increase linearly with $N$, $R^2_N \propto \langle m \rangle \propto N$.
For $b<b_c$ but close to $b_c$ they first show the same behaviour,
indicating that short polymers would be stretched at this $b$, but then
cross over to the collapsed phase in a very narrow range of $N$.

We claim that this is a first (yet inconclusive) indication for the first
order nature of the transition. More direct indications are obtained by
studying the histograms of $x$ and $m$. The histograms shown in the 
following figures are obtained by combining data from runs with different 
$q$ and $b$ and reweighting them. Combining MC results from different 
temperatures is not trivial for conventional Metropolis-type Monte Carlo 
algorithms where absolute normalization is unknown \cite{ferrenberg}. In
contrast, it is straightforward for PERM, since PERM gives directly
estimates of the partition sum and of the properly normalized histograms 
\be
   P_{q,b}(m,x) = \sum_{walks}q^{m'}\, b^{x'}\,\delta_{m,m'}\delta_{x,x'} \;.
\ee
Reweighting histograms obtained with runs performed nominally at $q$ and $b$
is trivially done by 
\be
   P_{q',b'}(m,x) = P_{q,b}(m,x) (q'/q)^m (b'/b)^x \;.
\ee
Combining results from different runs can then be either done by selecting for 
each $(m,x)$ just the run which produced the least noisy data (which was 
done here in most cases), or by assuming that the statistical weights of different runs
are proportional to the number of `tours' \cite{g97} which contributed to 
$P_{q,b}(m,x)$. 

In the present work we studied only single variable histograms $P_q(m)$
and $P_b(x)$, for which the above holds with the appropriate modifications.

Histograms of $m$ for fixed $q$ and $b$,
and for three different values of $N$, are shown in Fig.~12. For 
$N = 1500$ and $N=2000$ we see two peaks, corresponding to the collapsed 
(right hand, $m\approx 0.9N$) and stretched (left hand, $m\approx 0.4N$) 
phases. For $N>2000$ all chains would be collapsed, while for $N\le 1000$ 
all chains are stretched, in agreement with Fig.~11. 

\begin{figure}
 \begin{center}
   \psfig{file=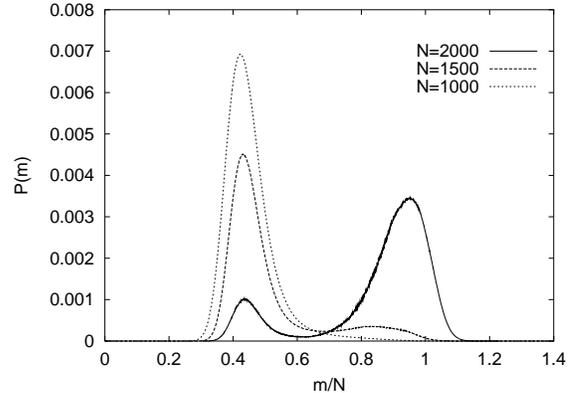,width=5.4cm, angle=270}
   \label{hm-3d}
   \caption{Histograms of the number of non-bonded nearest-neighbour pairs 
     $P(m)$ versus $m/N$ for $d=3,\; q=1.5,$ and $b=1.55<b_c$.
     The peaks near $m/N \approx 0.4$ correspond to the stretched phase, 
     the ones near $m/N\approx 0.9$ to the collapsed phase. Notice that chains
     with $N=1000$ are for these values of $b$ and $q$ entirely in the stretched
     phase (there is no peak near $m/N\approx 0.9$), in agreement with 
     Fig.~11. Normalization is arbitrary.}
   \end{center}
\end{figure}

\vglue -7mm
\begin{figure}
 \begin{center}
   \psfig{file=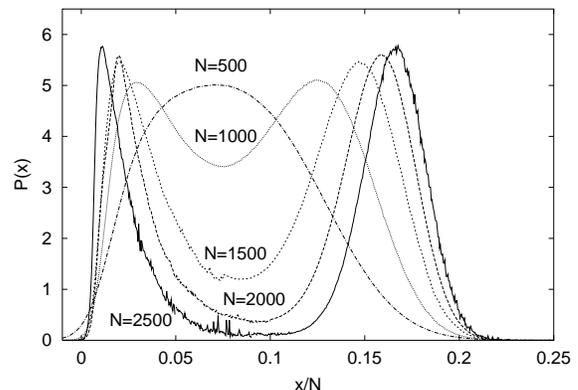,width=5.4cm, angle=270}
   \label{hx-3d}
   \caption{Histograms of the end point distance $P(x)$ versus $x/N$
     for $q=1.5$. Biases were adjusted so that both peaks have
     equal height: $b=1.4040\; (N=500)$, $1.4925\; (N=1000)$,
     $1.5386\; (N=1500)$, $1.5658\; (N=2000)$, $1.5855\; (N=2500)$.
     Normalization is arbitrary. The peak at $x/N \approx 0$ corresponds
     to the collapsed phase, the other to the stretched phase.}
  \end{center}
\end{figure}
\vglue -4mm

Analogous histograms of $x$ are shown in Fig.~13 for a
wider range of $N$. We now see an even more pronounced double peak
structure, with the left (right) peaks corresponding to the collapsed
(stretched) phase. In this figure we kept $q=1.5$ fixed but varied
$b$, so that both peaks have the same height for each 
$N$. In addition we adjusted the normalization arbitrarily such that 
all peaks have similar heights. We see clearly that the height of 
the minimum between the peaks shrinks to zero for $N\to\infty$, and 
that the horizontal distance between the peaks increases with $N$. 
Taken together, they form a clear indication for a first order transition. 
Notice that a double peak structure with decreasing minimum alone would  
not be a conclusive proof, as shown by the $\Theta$-point in 
dimensions $d\ge 4$ \cite{po00}.

The values $b=b_c(N)$ for which the two peaks to have equal height 
(indicated in the caption of Fig.~13) are effective finite
$N$ transition points. According to our phenomenological model we 
expect them to scale as
\be
     \mid b_c(N) - b_c \mid \propto N^{-1/3}\, .   \label{bc}
\ee
The values obtained from Fig.~13 are plotted in Fig.~14
against $N^{-1/3}$. From Eq.~(\ref{bc}) we expect them to fall onto
a straight line. This is indeed the case, and the extrapolation to 
$N\to\infty$ gives $b_c = 1.84\pm 0.04$.

\begin{figure}
 \begin{center}
   \psfig{file=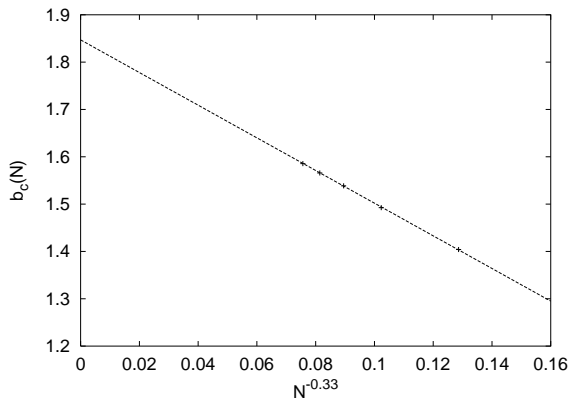,width=5.4cm, angle=270}
   \label{bc-3d}
   \caption{Effective transition points $b_c(N)$ versus $N^{-1/3}$ for $d=3,\;
      q=1.5$. The true transition point $b_c=1.84\pm 0.04$ is determined by 
      extrapolating to the $y$-axis.}
  \end{center}
\end{figure}

This estimate of $b_c$ is already more precise than any estimate we 
could obtain from Fig.~11. But an even more precise estimate 
is obtained by analyzing the partition sum itself. In Fig.~15 we
show $\log Z_N(q,b)+\mu_\infty N$ versus $N$. For small $b$ the curves
are close to the curve for $b=1$. As $b$ increases, the initial
(small-$N$) parts of these curves are straight lines with less and
less negative slopes. In this regime the polymer is stretched. As long
as these slopes are negative, the straight lines will intersect
the curve for $b=1$ at some finite value of $N$, say $N_c(b)$. Obviously
these are the chain lengths where $b = b_c(N_c)$ (see Fig.~14).

For $N>N_c(b)$, the values of $\log Z_N(q,b)+\mu_\infty N$ must deviate
from the straight lines. Indeed, the curves in Fig.~15 cannot
cross each other since \cite{mg01}
\begin{eqnarray}
  \log Z_N(q,b) & - & \log Z_N(q,b=1) \nonumber \\
     & = & \log [\langle \cosh(x\log b) \rangle_{b=1}]\nonumber \\
     & \geq & \log[1+\frac{(\log b)^2}{2} \langle x^2\rangle_{b=1}]\; . \label{znb1}
\end{eqnarray}
In the simulations, such a crossing can of course happen due to metastability
of the stretched phase. With our algorithm, the collapsed state has much
lower energy close to the transition point, but also much lower entropy,
so that it can easily be missed during a run with finite CPU time.
Whenever this happened, were plotted in Fig.~15 the value 
predicted by Eq.~(\ref{znb1}) instead of using the direct estimate of 
$Z_N(q,b)$.

Since the curve for $b=1$ becomes horizontal for $N\to\infty$,
the true phase transition occurs at that value of $b$ for which the
straight line in Fig.~15 is also horizontal. This can be estimated
very easily and with high precision, giving for $q=1.5$ our final estimate
$b_c \approx 1.856(1)$. This is in perfect agreement with the above
finite-$N$ extrapolation. Its error is dominated by
the uncertainty of $\mu_\infty$.

\begin{figure}
 \begin{center}
   \psfig{file=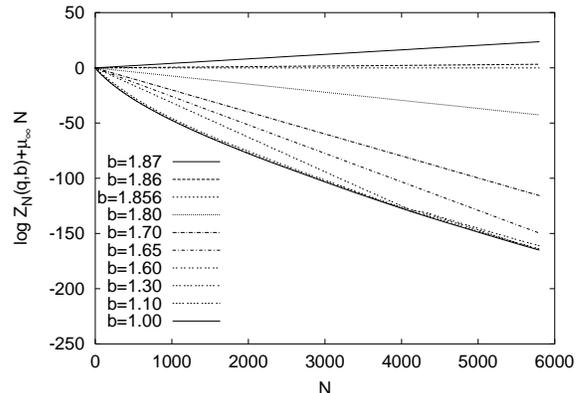,width=5.4cm, angle=270}
   \label{zg-3d}
   \caption{$\ln Z_N(q,b) + \mu_\infty N$ for $d=3, q=1.5$, and for various
      values of $b$. The value $\mu_\infty=-1.7530\pm 0.0003$ used in this plot
      was obtained from dense limit simulations on finite lattices.}
  \end{center}
\end{figure}

The results of $\log(b_c(q))$ and $\tilde{\sigma}(q)$ for a wide range of $q$
values are shown in Fig.~16. We see that both curves are roughly linear
near the $\Theta$ point, with slopes close to one. Exact scaling laws
$\log(b_c(q))\sim q-q_\theta$ and $\tilde{\sigma}(q)\sim q-q_\theta$ cannot 
be expected because of the strong logarithmic corrections at the $\Theta$ point
\cite{gh95,g97}. 

\begin{figure}
 \begin{center}
 \psfrag{log( b_c )}{\small $\log({\mathsf b}_{\mathsf c})$}
   \psfrag{(ss)}{\small $\tilde{\sigma}$}
   \psfrag{( q-q_t )/q_t}{\small ${{{\mathsf (q-q_\theta)/ q}}_{\theta}}$}
   \psfig{file=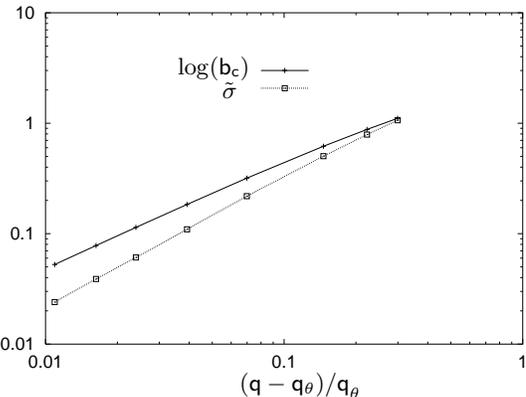,width=5.4cm, angle=270}
   \label{bceta-3d}
   \caption{Log-log plots of $\ln b_c(q)$ and of surface tensions 
      $\tilde{\sigma}(q)$ versus $(q-q_\theta)/q_\theta$ for $d=3$. The
      continuous lines are drawn to guide the eye.}
  \end{center}
\end{figure}

\section{Conclusion}

By employing the PERM algorithm to the BISAW model with
attractive interaction on square and simple cubic lattices,
we have studied the process
of stretching collapsed polymers in two and three
dimensions. In $d=2$ we find a second order transition,
in contrast to previous results for the bond fluctuation model
\cite{carmesin}. We do not know whether this is due to an inherent
difference in the models.
In $d=3$ a clear first-order phase transition is
observed, in agreement with all previous studies. But
it seems that the present study is the first with a detailed
study of the transition region and of finite size effects.

We indeed found the latter to be extremely important, in
particular for $d=3$. Estimates of the critical
force obtained without careful extrapolation
to $N\to\infty$ would be grossly wrong. This is
similar to the $\Theta$ collapse of unstretched polymers in 
high dimensions \cite{po00} where these finite size effects
even mimicked a first order transition, while the true transition 
is second order. We believe that we can exclude the latter for 
the present case. 

The most precise estimates of critical forces were obtained
via a direct comparison of the free energies in the two phases.
This is easily done with PERM, in contrast to most other
Monte Carlo methods. We found that the partition sum of the
collapsed phase is very closely related to that of
completely unstretched collapsed polymers. An essential part
of our numerical effort went indeed into improved estimates
of the latter. In particular, we verified that the concept
of a surface tension applies both to 2-d and 3-d collapsed polymers,
and for $d=3$ we verified its scaling with temperature.

We compared the finite $N$ corrections in detail with
phenomenological models which yield different order transitions
in $d=2$ and $d=3$, and found perfect agreement. The latter is a
bit surprising since fluctuations and surface layer thicknesses
are neglected in the latter. These approximations should become
exact in the limit of chain length $N\to\infty$, but our present
chains, with $N\approx 10^3 - 10^4$, could be expected a priori to
be much too short for this.

\section*{Appendix}

PERM is a particular depth-first implementation of sequential
importance sampling with reweighting \cite{liu}. Polymer chains are built
like random walks by adding one monomer at each step.

As in any such
algorithm we have the freedom to sample these steps from a wide
range of possible distributions, provided this additional bias is
taken into account by suitable weight factors. First of all, we used
a Rosenbluth like bias in avoiding steps which would lead to
self intersections. Among the other possible steps we selected those
parallel, antiparallel, and transverse to $\bf F$ with probabilities
$p_{+x}:p_{-x}:p_{\perp} = \sqrt{b}:\sqrt{1/b}:1$. The Boltzmann
factors for the pair interactions were taken into account entirely
by the weight factors and did not enter into the step probabilities.
Let us define
$p_{\perp}^{(0)}=1,\; p_{\pm}^{(0)} = b^{\pm1/2}$. Then we
have $p_i = 0$ if step $i$ is forbidden, and
\be
   p_i = {p_i^{(0)} \over \sum_{{\rm allowed}\;j} p_j^{(0)}}
\ee
else.
The corresponding weight factors are then
\be
   w_i = \frac{q^{m_n} b^{\Delta x_i}}{p_i} \qquad (\Delta x_i=0,\;1,\;{\rm or}\; -1) \, ,
\ee
where $m_n$ is the number of neighbours of the new site already
occupied by non-bonded monomers.

The total weight of a chain of length $n$ is then
$W_n = \prod_{n'\leq n} w_{i_{n'}}$.
Every time an $n$-th monomer is added to the chain, we update the
current estimate of the partition sum to
\be
   {\hat Z}_n = M^{-1}_n \sum_{\alpha = 1}^{M_n} W_n(\alpha) \, ,
\ee
where $M_n$ is the number of chains reaching length $n$ and
$W_n(\alpha)$ is the weight of the $\alpha$th chain. Chains are
cloned and pruned if their weight is above $3{\hat Z}_n$ and
below ${\hat Z}_n/3$, respectively.

For different sets of simulations we measured different 
observables. In the dense limit simulations, e.g., we measured only 
the partition sum, while the largest number of observables was 
measured for stretched chains on infinite lattices. There
we measured the partition sum $Z_n$, the average
end-to-end displacement $\langle x \rangle$ parallel to the
force, the average squared end-to-end distance $<R^2>$, and the number
of contacts between non-bonded monomers $<m>$ which
is a measure for the internal energy. We also measured histograms of the
parallel displacement and of the contact number. In addition  
during all runs we made also some technical control measurements. Most 
importantly, we generated histograms of {\it tour weight} distributions
\cite{tourweights} in order to test whether the results are statistically
reliable or not.

In all cases we used a single integer to label lattice sites, and 
used helical boundary conditions. If the lattice size was a power
of $2^d$, say $2^{kd}$ with integer $k$, then the neighbours of site
$i$ are $i\pm 1, i\pm 2^k,\ldots i\pm 2^{(d-1)k}$, and integers 
outside the range $[0, \ldots 2^{kd}-1]$ are brought back to this 
interval by means of bitwise AND with $2^{kd}-1$. If the lattice 
size was a different power of 2, say $2^p$, the numbers
$1, 2^k,\ldots 2^{(d-1)k}$ are replaced by the integers nearest to 
$1, 2^{p/d},\ldots 2^{(d-1)p/d}$.

In $d=2$ and for finite lattices, self avoidance and contacts were 
simply checked by means of bit maps: in an array of characters, each 
occupied site was marked `1', while each empty site was marked 
`0'. For opens systems in $d=3$ this would have needed too much 
storage and we used hashing. In this way we could implement 
effectively infinite lattices with relatively small computer 
memory.

Acknowledgements: We thank Dr. Walter Nadler for very helpful 
discussions.

\end{multicols}


\begin{thebibliography}{99}
\bibitem{footnote} Notice that we describe here polymers in
equilibrium. A real polymer, when stretched from a random
globule state, is not in equilibrium since knots cannot
unfold during and after the stretching, unless we deal with
phantom polymers.
\end{thebibliography}
\end{document}